\documentclass[onecolumn]{aastex631}

\shorttitle{B335 outflow}
\shortauthors{Hodapp et al.}

\begin{document}
\turnoffedit
\title{The outflow of the protostar in B335: I}

\correspondingauthor{Klaus Hodapp}
\email{hodapp@ifa.hawaii.edu}

\author{Klaus W. Hodapp}
\affil{University of Hawaii, Institute for Astronomy, 640 N. Aohoku Place, Hilo, HI 96720, USA}

\author{Laurie L. Chu}
\affil{IPAC, California Institute of Technology, 1200 E California Boulevard, Pasadena, CA 91125}

\author{Thomas Greene}
\affil{NASA Ames Research Center Space Science and Astrobiology Division M.S. 245-6 Moffett Field, CA 94035, USA}

\author{Michael R. Meyer}
\affil{Department of Astronomy, University of Michigan, 1085 S. University, Ann Arbor, MI 48109, USA}

\author{Doug Johnstone}
\affil{NRC Herzberg Astronomy and Astrophysics, 5071 West Saanich Rd, Victoria, BC, V9E 2E7, Canada}
\affil{Department of Physics and Astronomy, University of Victoria, Victoria, BC, V8P 5C2, Canada}

\author{Marcia J. Rieke}
\affil{Steward Observatory, University of Arizona, Tucson, AZ 85721, USA}
\author{John Stansberry}
\affil{STScI, Steven Muller Building, 3700 San Martin Drive, Baltimore, MD 21218, USA}
\author{Martha Boyer}
\affil{STScI, Steven Muller Building, 3700 San Martin Drive, Baltimore, MD 21218, USA}
\author{Charles Beichman}
\affil{IPAC, California Institute of Technology, 1200 E California Boulevard, Pasadena, CA 91125}
\author{Scott Horner}
\affil{NASA Ames Research Center Space Science and Astrobiology Division M.S. 245-6 Moffett Field, CA 94035, USA}
\author{Tom Roellig}
\affil{NASA Ames Research Center - Ames Center for Exoplanet Studies (ACES), CA, USA}
\author{George Rieke}
\affil{Steward Observatory, University of Arizona, Tucson, AZ 85721, USA}
\author{Eric T. Young}
\affil{Universities Space Research Association, 425 3rd Street SW, Suite 950, Washington DC 20024}



\begin{abstract}

The isolated globule B335 contains a single, 
low luminosity Class 0 protostar associated with a bipolar nebula 
and outflow system seen nearly perpendicular to its axis. 
We observed the innermost regions of this outflow 
as part of JWST/NIRCam GTO program 1187, 
primarily intended for wide-field slitless spectroscopy of background stars behind the globule. 
We find a system of expanding shock fronts with kinematic ages of only a few decades emerging 
symmetrically 
from the position of the embedded protostar, which is not directly 
detected at NIRCam wavelengths. 
The innermost and youngest of the shock fronts studied here shows strong emission from CO.
The next older shock front shows less CO and the third shock front shows only H$_2$ emission in our data.
This third and most distant of these inner shock fronts shows substantial
evolution of its shape since it was last observed with high spatial resolution in 1996 with Keck/NIRC.
This may be evidence of a faster internal shock catching up with a slower one and of the two shocks merging.

\end{abstract}





\section{Introduction}

Stars form out of dense cores of molecular gas by accretion of material.
In almost all circumstances, the gas accreted in this process has
much larger angular momentum than the newly formed star will have. As a way
of shedding the excess angular momentum, mass accretion onto a protostar
or young star is associated with outflow activity.
For the youngest stars in their main accretion phase, the outflow is
in the form of a collimated jet with internal shock fronts radiating
primarily in the emission lines of H$_2$. This fast jet interacts turbulently
with the ambient molecular material and entrains it into a broader, slower
molecular outflow that is best studied by radio emission of CO.
In later phases of mass accretion onto almost completely formed young
stellar objects (YSO), well collimated, very high velocity jets radiating
in atomic emission lines are commonly found.
Outflow and jet phenomena in young stars were recently reviewed by
\citet{Bally.2016.ARAA.54.491B}
and we refer to the references therein for an outline of the history
of this field.

The isolated globule Barnard 335 in Aquila
\citep{Barnard.1927.pasr.book.B}
was observed as part of JWST program 1187, 
primarily to map the broadband extinction and to measure 
ice absorption column densities with NIRCam Wide Field Slitless Spectroscopy (WFSS).
Details of our sample selection and preliminary ground-based results
on the onset of ice formation in those cores was presented by
\citet{Chu.2021.ApJ.918.2.extinction.maps}.
B335 is the only molecular core in our sample of three cores for program 1187 that 
has an embedded protostar, discovered by
\citet{Keene.1983.ApJ.274L.43.B335.protostar}.

The distance to B335 was first measured by 
\citet{Tomita.1979.PASJ.31.407.distance}
at 250 pc 
and this was long used as the canonical distance. 
Using similar purely photometric methods, 
\citet{Olofsson.2009.AA.498.455O.B335.distance}
measured a distance in the range of 90 - 120 pc. 
They had already mentioned a possible association of the B335 cloud 
with the star HD 184982 on the basis that this star appears to be illuminating 
the cloud in addition to the general interstellar radiation field. 
\citet{Watson.2020.RNAAS.4.88.B335.distance}
obtained new images of this reflection nebulosity 
and used the Gaia DR2 distance to HD 184982 of 164.5 pc as a proxy 
for the distance of the B335 molecular core. 
For this paper, we adopt this distance value  
of 164.5 pc.

The larger molecular cloud surrounding the Bok globule
B335 was found by
\citet{Frerking.1982.ApJ.256.523.B335.protostar}.
The molecular outflow associated with the embedded
protostar was discovered and mapped by
\citet{Hirano.1988.ApJ.327L.69.B335.inclination}
who
also determined that the outflow axis is inclined only
10$\arcdeg$ from the plane of the sky, i.e., the bipolar
outflow is seen nearly perpendicular to its axis.
The molecular core B335 is the densest part of a larger
cometary globule, suggesting that an external wind shaped
the globule and triggered the star formation in it,
as was discussed in the case of the globules associated
with the Gum nebula by 
\citet{Reipurth.1983.AA.117.183.Gum.neb.globules}.
The pattern of background star polarization, first mapped by
\citet{Hodapp.1987.ApJ.319.842.B335.polarization} is also
consistent with the globule having been formed by an
external wind.

The bipolar outflow emerging from the single protostar
in B335 was studied by
\citet{Stutz.2008.ApJ.687.389.B335.Spitzer}
based on Spitzer Space Telescope imaging data.
Out to a wavelength of 8 $\mu$m the object appears as a 
bipolar nebula. Only at wavelengths of 24 $\mu$m and longer
does the object appear as an unresolved point source,
probably direct light from 
the central embedded protostar.
The ALMA-based studies of B335 by
\citet{Evans.2015.ApJ..814.22.position} and \citet{Evans.2023.ApJ.943.90.B335.variability} found clear
evidence for infall from the redshifted absorption seen
in HCN and HCO$^+$, estimated the age of the accreting
protostar at 5.0 $\times$ 10$^4$ yrs, and found evidence
for variability and episodic accretion. Based on ALMA data
\citet{Bjerkeli.2019.AA.631A.64.ejection.bullet} found evidence for the
recent (around 2015) ejection of a molecular bullet and
\citet{Bjerkeli.2023.AA.677A.62.infall} presented evidence for infall
faster than free-fall velocity, interpreted as evidence for
variable infall and accretion.
WISE W2 (4.6 $\mu$m)
scattered light from the embedded protostar showed an outburst between
2015 and 2022 \citep{Kim.2023.arXiv231205781K}, again supporting variable
accretion.

Optical Herbig-Haro shock fronts emerging from the B335 globule 
were found first by 
\citet{Vrba.1986.AJ.92.633.HH119}
and then confirmed and added to by 
\citet{Reipurth.1992.AA.256.225.B335.HH}
who labelled them as HH 119 A-C. 
In the infrared, the first major imaging study was done by 
\citet{Hodapp.1998.ApJ.500L.183.B335}
who labelled shock fronts found in the H$_2 1–0$ S$(1)$ emission line, 
but not confirmed by optical spectroscopy as HH 119 IR1-5.
This naming scheme was expanded to include newly identified shocks by \citet{Galfalk.2007.A&A.475.281.B335.outflows}.
The strong extinction by the nearly edge-on disk and the dense inner envelope around the protostar in B335 
has prevented any studies of infrared shock fronts in the immediate vicinity of the protostar prior to JWST.

Our JWST observing program 1187 on B335 was cut short by a guide star acquisition
failure on the second of two different telescope pointings. The data presented here are
about half of the originally intended data set, which explains the incomplete spatial coverage. 

In this paper, we present our imaging and wide-field slitless spectroscopy (WFSS) data
in section 2. We discuss the outflow cavity in section 3.1,
the position and proper motion of the protostar in section 3.2,
the shock proper motions in 3.3, and the emission spectrum of some of
the shocks in section 3.4.

\newpage

\section{Observations and Results}

The results reported here are based on guaranteed time observations (GTO)
with the James Webb Space Telescope (JWST) 
\citep{Gardner.2023.PASP.135f8001.JWST.mission}.
The imaging observations were obtained as part of the 
JWST Near-Infrared Camera (NIRCam) 
\citep{Rieke.2023.PASP.135b8001.NIRCam}
Wide Field Slitless Spectroscopy (WFSS) 
of background stars behind B335 aimed at mapping the 
column density of H$_2$O, CO$_2$, and CO ice (program 1187). 
The use of the NIRCam grisms for slitless spectroscopy was described by 
\citet{Greene.2017.JATIS.3c5001.grism}.
The observations were carried out on 2023 April 21 UTC (MJD 60059). 
WFSS data were obtained 
through the four medium width filters
F410M, F430M, F460M, and F480M, whose bandpasses cover the CO$_2$ and CO
ice features  
The DOI 
\dataset[10.17909/p188-aj80]{\doi{10.17909/p188-aj80}}
contains all the spectra extracted by the stage 3 pipeline, which
were not used in this study. However,
the WFSS data were used in stage 2 calibration form and
the filenames used are listed in Table 1. They can be filtered from
the dataset pointed to by the DOI. 
Direct images in the F277W, F300M, F356W, and F444W filters were obtained 
as part of the source registration and wavelength calibration procedure for the slitless spectroscopy
and these data is available at MAST: \dataset[10.17909/qa3f-5703]{\doi{10.17909/qa3f-5703}}. 

Due to the strong extinction near the B335 protostar, 
only the long-wavelength module data at wavelengths longer than 2.5 $\mu$m
proved useful for this paper, and are
summarized in Table 1.
The imaging data were processed by STScI using the JWST Science Calibration Pipeline
(\dataset[10.5281/zenodo.8067394]{\doi{10.5281/zenodo.8067394}})
version 1.9.6 and 
calibration file $jwst\_1084.pmap$ with astrometric calibration based on the Gaia DR3 catalog.
Figure 1 shows an overview color image of B335, composed of the
images in the F277W (blue channel), F356W (green channel), and F444W (red channel), in logarithmic scaling to emphasize low flux regions.

\begin{deluxetable*}{llrcccl}[hbt!]
\tabletypesize{\scriptsize}
\tablecaption{Observations Details\\}
\tablewidth{0pt}
\vspace{0.5cm}
\tablehead{
\colhead{Filter} & \colhead{grism} & \colhead{Total Integration} & \colhead{NGROUPS} &
\colhead{NINTS} & \colhead{Date [MJD]} & \colhead{MAST File Name}}
\startdata
F277W  & clear & 773 & 4 & 2 & 60055 & jw01187-o011\_t001\_nircam\_clear-f277w\_i2d.fits \\
F356W  & clear & 344 & 4 & 2 & 60059 & jw01187-o014\_t001\_nircam\_clear-f356w\_i2d.fits \\
F444W  & clear & 1374 & 4 & 2 & 60059 & jw01187-o015\_t001\_nircam\_clear-f444w\_i2d.fits \\
F410M  & GRISMC & 172 & 8 & 1 & 60059 & jw01187015002\_2101\_0001\_nrcalong\_rate.fits \\
F410M  & GRISMC & 172 & 8 & 1 & 60059 & jw01187015002\_210c\_0001\_nrcalong\_rate.fits \\
F430M  & GRISMC & 172 & 8 & 1 & 60059 & jw01187015002\_2103\_0001\_nrcalong\_rate.fits \\
F430M  & GRISMC & 172 & 8 & 1 & 60059 & jw01187015002\_210e\_0001\_nrcalong\_rate.fits \\
F460M  & GRISMC & 172 & 8 & 1 & 60059 & jw01187015002\_2105\_0001\_nrcalong\_rate.fits \\
F460M  & GRISMC & 172 & 8 & 1 & 60059 & jw01187015002\_210g\_0001\_nrcalong\_rate.fits \\
F480M  & GRISMC & 172 & 8 & 1 & 60059 & jw01187015002\_2107\_0001\_nrcalong\_rate.fits \\
F480M  & GRISMC & 172 & 8 & 1 & 60059 & jw01187015002\_210i\_0001\_nrcalong\_rate.fits \\
\enddata
\end{deluxetable*}

\newpage


\begin{figure*}[hbt!]
    \begin{center}
	\includegraphics[angle=0.,scale=0.5]{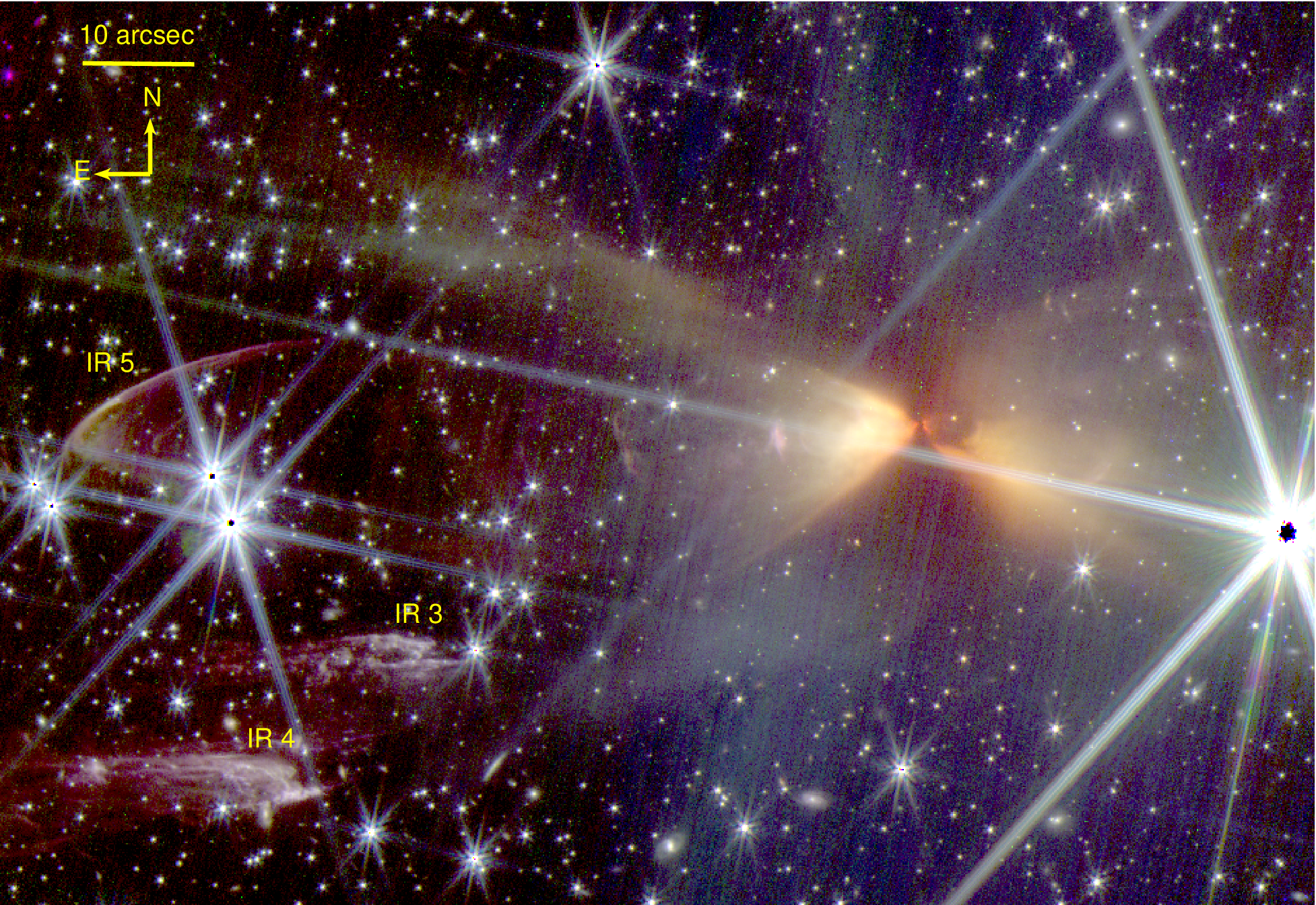}
	\caption{RGB color composite of the central 118$\arcsec \times$ 81$\arcsec$ of the B335 outflow in canonical orientation with North up and
 East left. The NIRCam filter F444W is red,
F356W is green, and F277W is blue, the flux scaling is logarithmic. Bright stars that saturated the detector pixels, 
most notably the bright background star 2MASS J19365867+0733595 at the western edge of the image, are displayed with black centers.
The previously known infrared shock fronts IR3 - IR5 \citep{Hodapp.1998.ApJ.500L.183.B335} are labelled.}
	\end{center}
\end{figure*}


\begin{figure*}[hbt!]
\begin{center}
	\includegraphics[angle=0.,scale=0.60]{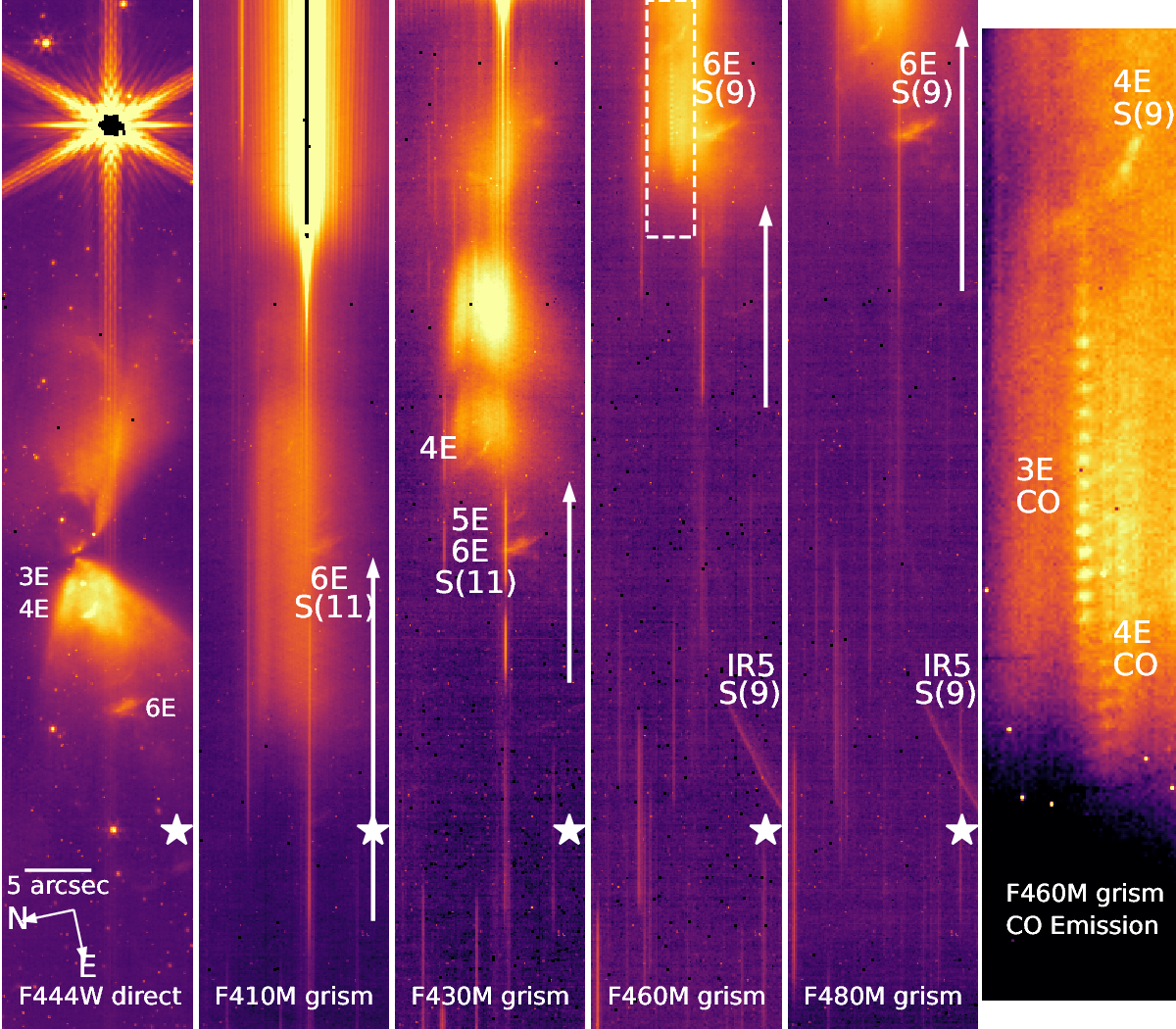}
	\caption{The left panel is a direct image in the F444W filter. The four middle panels are 
 cutouts of the WFSS images through the F410M, F430M, F460M, and F480M filters. The right panel is
 a magnified cutout of the F430M grism image showing the CO emission from shock fronts 3E and 4E.
All images are in the detector coordinate system, the grism (GRISMC) dispersion direction is vertical.
Each filter transmits a wavelength
range of the dispersed image of the scene, with spatial and spectral information intermixed.
In each of central four grism images, we indicate the position of one star with an asterix symbol $\approx$ 5\arcsec to
the right on the side of the frame, and indicate the transmission region of each filter (50\% point) and
the direction of increasing wavelengths. In the spectrum of the star indicated by this symbol, 
absorption by CO$_2$ ice is clearly seen in the F430M filter, and absorption by CO is seen in the
F460M and F480M images. We have labelled the positions of the images of shock fronts 4E, 5E, and 6E
in the H$_2$ 0-0 S(11) line in the F410M and F430M panels, and in the S(9) line in the F460M and F480M panels.
In the F460M panel, CO emission from shock 3E and some from shock 4E are seen, and shown magnified in
the right panel for clarity. The very bright star near the top of the images is saturated in the
direct F444W image and the F410M WFSS image.
}
	\end{center}
\end{figure*}

For the WFSS observations, a grism, in our
case the GRISMC in the NIRCam pupil wheel, disperses the light from every
object in the field into a spectrum. Individual medium band filters were
used to record sections of this spectrum. When using medium bandpass filters,
the spectrum section in each of the WFSS images only 200 - 300 pixels long,
so that the problem of spectra of different objects
overlapping is mitigated. As it turned out, without this sub-division of the spectral range, 
the CO emission from the bright compact shock front would have
been masked by flux from the reflection nebula and bright stars.
Figure 2 shows cutouts of the inner regions of the B335 bipolar nebula in the orientation of the NIRCam detectors, where the dispersion of the GRISMC is along the column axis. The left panel is the direct F444W image for orientation. The other panels show the same range of columns and a wavelength
dependent shift in the vertical (dispersion) direction.

The WFSS spectra of the extended shock fronts were extracted with custom python routines from the
the flux calibrated ``rate'' files produced by the STScI pipeline.
The wavelength
calibration was done using the H$_2$ emission lines in the object spectra
themselves. 
Slitless spectroscopy data in general intermix spatial and spectral
information and the spectral resolution depends on 
the spatial extent of the object. Since the
shock fronts studied here are extended and of complex morphology,
our data are suitable for detecting the presence of emission lines,
but not for studies of the radial velocity or its line structure.
In extracted spectra, the lines for one object are
recorded at the proper wavelengths, but other objects, spatially
offset, will also contribute to the spectra and their lines are
recorded at correspondingly offset wavelengths. 

We had two visits scheduled for WFSS observations through the four medium width filters.
Visit 15:1, covering the western part of the globule, was
not successful due to guide star acquisition problems, and a repeat 
request was approved.
This will give more imaging and spectroscopy in the western part of the globule, 
and hopefully allow better proper motion studies. 
The repeat observations, now labelled visit 65:1, are expected to be
scheduled in April or May of 2024, and will be published separately.

\section {Discussion}

\subsection{The outflow cavity}
The central region of the B335 outflow appears as a bipolar nebula
in our images. 
Other studies have shown that we see this outflow nearly edge-on 
and that the extinction towards the protostar must be extremely high. 
\citet{Chandler.1990.MNRAS.243.330.B335.embedded}
have measured an A$_V$ $>$ 320 mag. and 
\citet{Keene.1983.ApJ.274L.43.B335.protostar}
have shown that the protostar is only detectable 
at far infrared and radio wavelengths. 
\citet{Stutz.2008.ApJ.687.389.B335.Spitzer}
demonstrated that the embedded protostar only becomes directly
visible at 24$\mu$m in Spitzer Space Telescope images.
 
Indeed, even our longest wavelength image, in the JWST F444W filter,
does not show the protostar itself.
Recently, \citet{Evans.2023.ApJ.943.90.B335.variability} and
\citet{Kim.2023.arXiv231205781K} pointed out that the
WISE/NEOWISE mission
\citep{Wright.2010.AJ.140.1868.WISE}, recorded a pronouced outburst of B335 in the W2 (4.6$\mu$m).
The outburst lasted about 12 years and is just now ending in 2023, with some uncertainty since there is only one data point to define the quiescent baseline.
Using a sophisticated radiative transfer model of the B335 protostar,
\citet{Evans.2023.ApJ.943.90.B335.variability}
have interpreted the W2 flux maximum as evidence for a temporary increase
in the protostar luminosity up to $\approx$ 22 L$_\odot$.
This near infrared variability coincides with an increase
in radio line flux observed in 2018 with ALMA by \citet{Evans.2023.ApJ.943.90.B335.variability} and is therefore indicative of an increase in accretion activity
on the B335 protostar.

Due to the r = 5\arcsec photometry aperture and the poor spatial resolution of the WISE telescope, 
these measurements covered essentially all the scattered light
from the bright parts of the B335 outflow cavities.
At the time of our $Spitzer$ observations, the B335 protostar had a luminosity of 16 L$_\odot$
\citep{Kim.2023.arXiv231205781K}, a factor of 4 above the 2010 quiescent value.

Archival $Spitzer Space Telescope$ Infrared Array Camera (IRAC) \citep{Fazio.2004.ApJS.154.10.Spitzer.IRAC} 
images
(\dataset[10.26131/IRSA413]{\doi{10.26131/IRSA413}})
from 2004 and images taken with the same instrument by \citet{Chu.2021.ApJ.918.2.extinction.maps} in 2016
during the outburst are shown in Fig.~3. In both IRAC channels 1 (3.55 $\mu$m) and 2 (4.49 $\mu$m), the bipolar nebula was substantially brighter
in 2016 than in 2004, confirming the brightness outburst. 
The Spitzer photometry shown by \citep{Kim.2023.arXiv231205781K} based on the 2004 data shows the object above
the quiescent brightness, but this data point was taken with an aperture of 12\arcsec, so is not directly
comparable to the WISE data points. For our discussion, the only important point is that the difference
images in Fig.~3 do not show the full amplitude of the outburst of the protostar.

With the better spatial resolution of $Spitzer$, the difference images (2016-2004)
show that the change in brightness was not entirely uniform across the eastern lobe of the bipolar reflection nebula, but
that changes in the local illumination conditions contributed to the brightness changes. Such spatial variations are common
in bipolar reflection nebulae, e.g., in Hubble's Variable Nebula NGC 2261 \citep{Hubble.1916.ApJ.44.190.var.neb}, and also
observed in several younger and more deeply embedded bipolar such as Cep A by \citet{Hodapp.2009.AJ.137.3501.CepA.var} and
L483 by \citet{Connelley.2009.AJ.137.3494.L483.var}. However, in B335, the latest outburst was dominated by an overall increase
in the protostar luminosity, and extinction and variations along the scattering light path are a minor contribution.
Our JWST imaging data were taken close to the end of this outburst, but with the luminosity still about 50\%
above the quiescent value of 4 L$_\odot$.

\begin{figure*}[hbt!]
\begin{center}
	\includegraphics[angle=0.,scale=0.6]{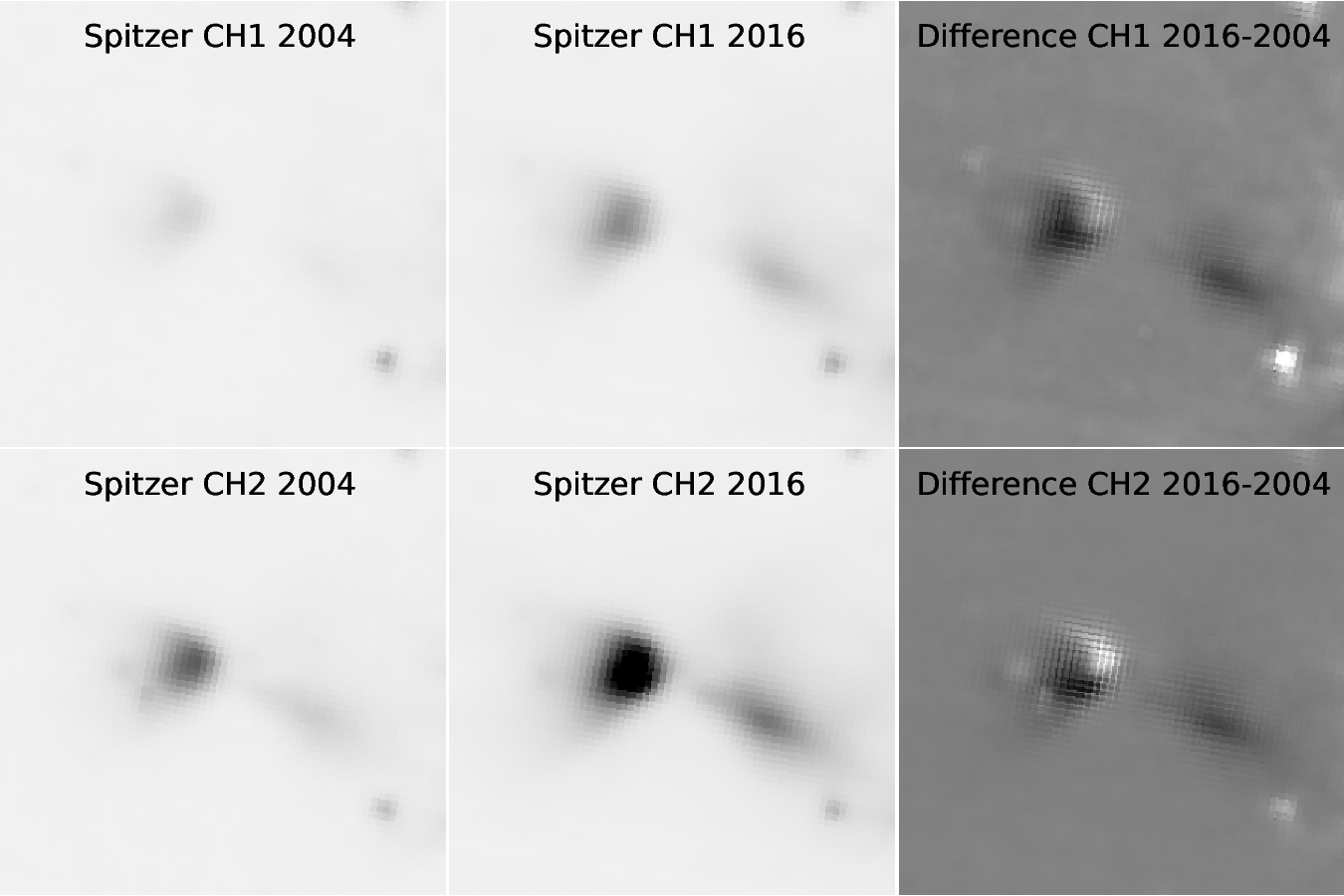}
	\caption{All panels are $40\arcsec \times 40\arcsec$ cutouts of Spitzer mosaic images. 
The top panels are IRAC images in channel 1 (3.55 $\mu$m)
and the lower row is in channel 2 (4.49 $\mu$m).
Each row shows the 2004 image in the left, the 2016 image in the center, and the difference, scale for average zero signal at right.
In addition to the overall change of brightness, changes in the spatial flux distribution are apparent.}
\end{center}
\end{figure*}

The eastern outflow cavity is outlined in the wide-field view of B335 
in the RGB composite in Fig.\ 1. 
The two lobes of the bipolar nebula are of different surface brightness, 
with the eastern lobe being brighter, 
indicating that this eastern lobe suffers less extinction 
and is on the front side of the equatorial disk around the protostar.
This essentially confirms the finding by
\citet{Hirano.1988.ApJ.327L.69.B335.inclination}
that the eastern outflow is slightly inclined towards the observer.
The outflow cavity walls, seen in CO velocity-integrated line emission
\citep{Kim.2023.arXiv231205781K}
or even better in the CS line \citep{Evans.2023.ApJ.943.90.B335.variability}
closely resemble the outflow cavity in our images (Fig. 1), including the
fact that the eastern outflow lobe appears narrower than the western lobe.

Beyond the outflow cavity that is illuminated by the embedded protostar, the
B335 molecular core exhibits coreshine \citep{Steinacker.2010.AA.511A.9S.coreshine},
i.e. its edges are outlined at infrared 
wavelength by the scattered interstellar radiation field. The central
area of the core appears darker than the edges in Fig.~1 and the dark
area is extended roughly in north-south direction, consistent with a large thick disk
seen edge-on.
This same phenomenon has been described as a
``shadow'' minimum in the extended flux
at 3.6~$\mu$m and 8.0~$\mu$m approximately 10\arcsec\ to the 
south of the protostar by
\citet{Stutz.2008.ApJ.687.389.B335.Spitzer} based on $Spitzer$ images.
The extended flux minimum is clearly associated with the nearly complete absence
of background stars seen in this area of the cloud.
This qualitatively confirms the interpretation by
\citet{Stutz.2008.ApJ.687.389.B335.Spitzer}
that this depression in the extended emission is
due to the flattened structure of the B335 molecular core.
A detailed mapping of the extinction in B335 was the
original purpose of our JWST observing program and will
be published separately from this paper on the outflow.

\subsection{Position of the embedded protostar}
The protostar in B335 has been observed at radio frequencies numerous times,
at different wavelengths and with basically every suitable telescope system available.
Since there is only one protostellar source in B335, the centroid of the
continuum flux distribution measures the position of the protostellar object
irrespective of wavelength.
For the purpose of determining the proper motion of the protostar, only
interferometric measurements have sufficient spatial resolution and
precision of the position.
In the lower panel of Fig.\ 4, we have indicated the position at 3.6 cm wavelength reported
by \citet{Reipurth.2002.AJ.124.1045.B335.VLA} and measured in 2001.
More recently, the best position measurements are from ALMA at 1.2 or 1.3 mm, and
we have included the measurements from
\citet{Maury.2018.MNRAS.477.2760.B335.ALMA} with epoch 2015,
\citet{Imai.2019.ApJ.873L.21.B335.ALMA} with epoch 2017
and that of
\citet{Okoda.2022.ApJ.935.136.ALMA.position}
with epoch 2019 in Fig. 4.
For clarity, it should be mentioned that these radio positions are measured against
the celestial reference coordinate system, and not relative to the B335 molecular cloud.
The protostar has always been located at the center of the bipolar nebula and plotting
those older radio positions on a contemporary image does not suggest otherwise.
Rather, the whole bipolar nebula, and likely, the whole B335 globule, has the
same proper motion as the protostar radio source.

\begin{figure*}[hbt!]
\begin{center}
	\includegraphics[angle=0.,scale=0.25]{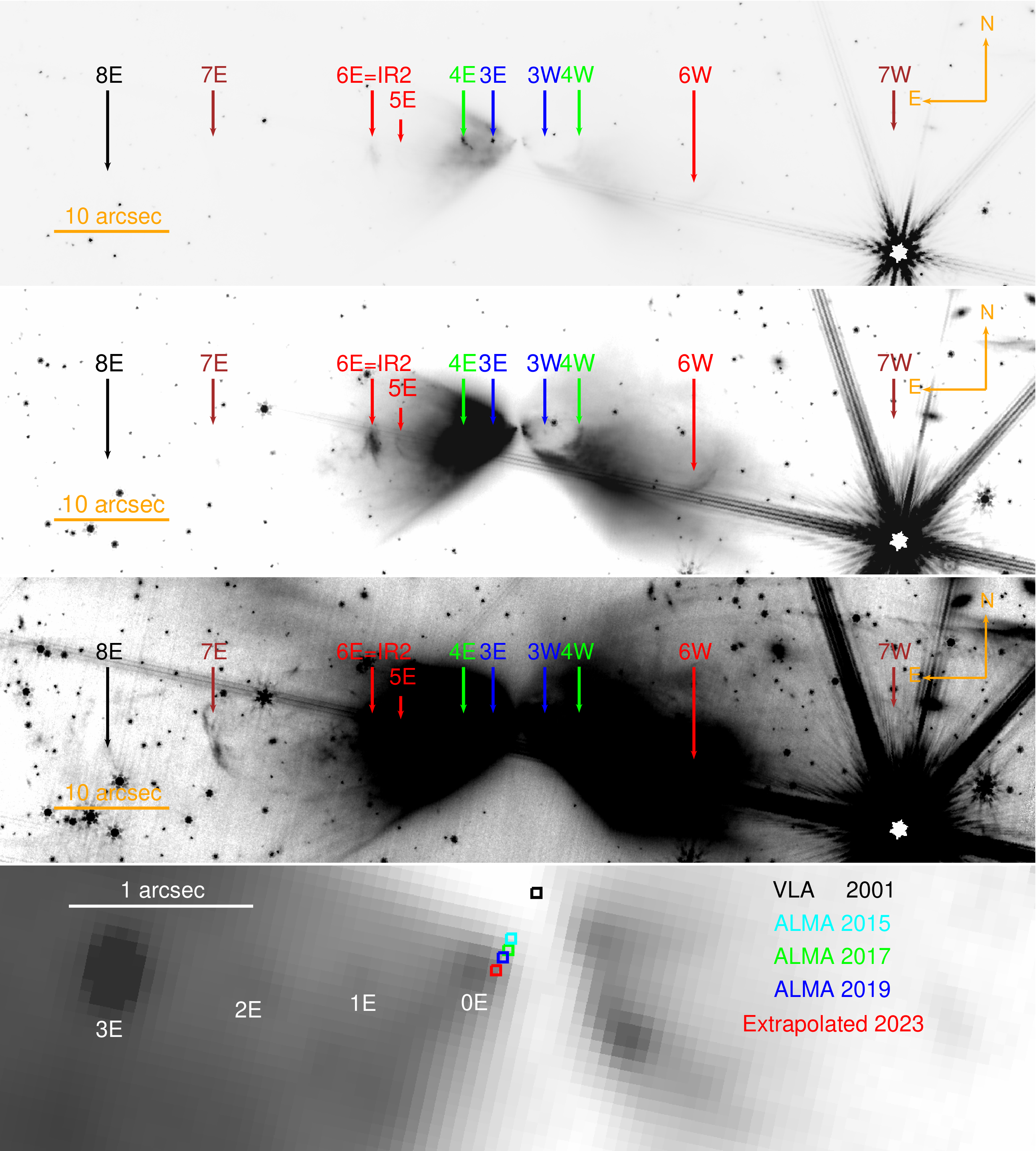}
	\caption{The top three panels are $90\arcsec \times 25\arcsec$ cutouts 
 of the F444W image in the canonical orientation with North up and East left. The top three panels are different
flux scalings to show all the symmetric shock front despite the difference in extinction towards the two outflow lobes.
The bottom panel is zoomed by a factor 16 into the central region near the protostar. The faint, innermost
features in the eastern outflow lobe are identified. 
We have also indicated the radio positions reported
by \citet{Reipurth.2002.AJ.124.1045.B335.VLA} with epoch 2001,
\citet{Maury.2018.MNRAS.477.2760.B335.ALMA} with epoch 2015,
\citet{Imai.2019.ApJ.873L.21.B335.ALMA} with epoch 2017
and that of
\citet{Okoda.2022.ApJ.935.136.ALMA.position}
with epoch 2019.
From those, we
have extrapolated the position of the protostar at the epoch of our JWST NIRcam image, which
is indicated by the red box in the bottom panel.}
\end{center}
\end{figure*}

The proper motion of the B335 protostar is 19 mas yr$^{-1}$ with an estimated uncertainty
of $\pm$ 1 mas yr$^{-1}$ based on the positions by \citet{Reipurth.2002.AJ.124.1045.B335.VLA} and
\citet{Okoda.2022.ApJ.935.136.ALMA.position}.
Extrapolating from the 
\citet{Reipurth.2002.AJ.124.1045.B335.VLA}
and
\citet{Okoda.2022.ApJ.935.136.ALMA.position}
measurements, we get the extrapolated position of the protostar at the epoch of
the JWST imaging observations as 19 37 0.9025 +7 34 09.42 (J2000.0) with an uncertainty of $\approx$ 20 mas 
and have indicated this by the red box
in Fig.\ 4.
This position places the protostar not in the center of the high-extinction
region separating the two outflow lobes, but close to its eastern edge,
consistent with the eastern lobe being oriented slightly towards
the observer, as was already noted by
\citet{Hirano.1988.ApJ.327L.69.B335.inclination} and discussed in the
subsection above. 
The protostar is located close to what appears to be
the first emission knot in the string of shock fronts, and that we
therefore label HH119 JWST 0E. At present, we do not have spectroscopic
data to prove that this is indeed a shock front, rather than scattered
continuum light. 

\newpage

\subsection{Shock proper motion}
Numerous outflow shock fronts are found within the outflow
cavity by
\citet{Vrba.1986.AJ.92.633.HH119},
\citet{Reipurth.1992.AA.256.225.B335.HH},
\citet{Hodapp.1998.ApJ.500L.183.B335}, and
\citet{Galfalk.2007.A&A.475.281.B335.outflows}.
The newly discovered shock fronts described here are named HH 119 JWST 
with a numbering scheme acknowledging the symmetric nature of these shocks:
for example, the closest of the newly discovered shock fronts east 
of the protostar is HH 119 JWST 0E, labelled 0E in Fig.\ 4. 
In this paper, in particular for labels in figures, we will abbreviate the
shock front names to just the last two characters, e.g., 0E.
Of these shock fronts, only HH 119 JWST 6E has previously been labelled as IR2.
Figure~1 shows three large shock fronts in the eastern half 
of the bipolar outflow and outside of the more detailed view in Fig.\ 4: IR3, 4 and 5 in the nomenclature of
\citet{Hodapp.1998.ApJ.500L.183.B335}. 

The proper motion measurements of 
\citet{Galfalk.2007.A&A.475.281.B335.outflows}
showed that shock front IR2, the only one of the inner shock fronts that was found
by ground-based observations \citep{Hodapp.1998.ApJ.500L.183.B335}, 
and D = IR5, the large thin bow shock front, 
have large proper motions around 200 mas yr$^{-1}$.
Both are located
on the central axis of the outflow cavity, defined by a string
of smaller shock fronts closer to the protostar, shown in Fig.~4. 
We will refer to these shocks as axial shocks.

In contrast, IR3 and IR4a are located farther away from that central axis and have an order of magnitude smaller proper motions
of around 20 mas yr$^{-1}$. We will refer to these and similar shocks as peripheral shocks.
This distinction is also seen in the morphology of these shocks: 
The axial shock IR5, in particular, is clearly a large shock front of the outflow 
from the protostar into ambient gas, with the bow of the shock front pointing away from the protostar.
In contrast, the bows of the peripheral shock fronts IR3 and IR4a appear 
to point towards the protostar, 
indicating a shock of the outflow on a stationary denser clump of material, 
with a wake of outflow gas flowing around the obstacle.
This morphology and their location away from the outflow axis suggest
that IR3 and IR4a may be shocks formed by the interaction of the
outflow with the cavity walls.

To generalize this finding: High proper motion shock fronts are found mostly along
a fairly narrow jet near the center of the outflow cavity,
while low proper motion shock fronts are less concentrated
and may be formed by interaction of the outflow with the
cavity walls.

To illustrate the proper motion of shock fronts,
we compare the JWST results with the $K$-band image of B335
obtained by
\citet{Hodapp.1998.ApJ.500L.183.B335}
on 1996 Oct.\ 1 (MJD = 50357  ) with the Keck I NIRC camera.
This image is the deepest ground-based image of B335 and it shows
part of the reflection nebulosity in the eastern outflow lobe,
and, in particular, the shock front named IR2. 
Fig.\ 5 shows cutouts of the 2023 JWST F277W image and
the 1996 Keck NIRC $K$-band image. The images were aligned to optimally match the
background stars within the limits imposed by their random proper motions. 
The rms shift of individual stars on the 1996 Keck images relative to their position in the 2023 JWST images was 221 mas or 8 mas yr$^{-1}$.

The axial shock front 6E = IR2, shows a dramatic position
shift and major morphological changes. 
In contrast, the peripheral shock front IR~3 (lower row in Fig. 5) that was just recorded in the south-eastern corner
of the $K$-band image, shows very little motion.

\begin{figure}[hbt!]
\begin{center}
 \includegraphics[angle=0.,scale=0.50]{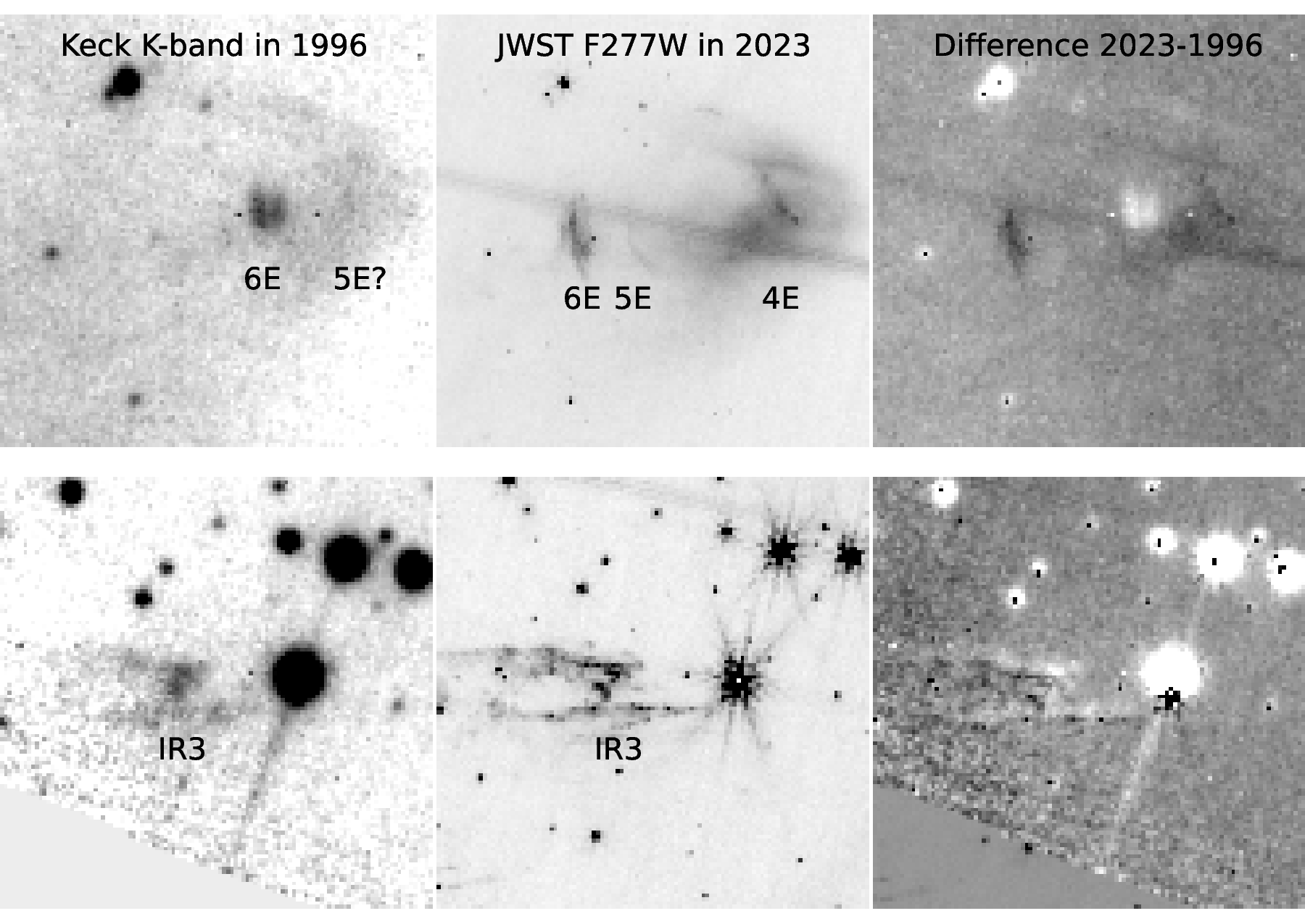}
 \caption{Comparison of shock position and morphology between 1996 (Keck NIRC $K$-band image) and 2023 (JWST NIRCam F277W image).
 The top row shows the 1996 K-band image of the inner shock fronts on the left, the 2023 F277W image in the center, and the difference
 on the right. Shock front 6E has moved substantially and has changed shape from two distinct shocks to one merged shock.
 The lower row shows the same for shock front IR~3 that shows minimal morphological change and only a very small (compared to 6E) change
 in position. The brightest star (anonymous) in this cutout shows high proper motion.
}
\end{center}
\end{figure}

The shock front 
6E = IR2 
was measured by
\citet{Galfalk.2007.A&A.475.281.B335.outflows}
to be 183 mas yr$^{-1}$.
We essentially confirm this proper motion with a measurement
of the proper motion from the epoch 1996 Keck image of
\citet{Hodapp.1998.ApJ.500L.183.B335} and our
JWST F277W image, the closest equivalent to the $K$-band 
in our data set. Our measurement of the proper motion
is 166 mas yr$^{-1}$ or 129.4 km\,s$^{-1}$ transverse velocity. 
The accuracy of this value is limited
by the morphological changes that this shock front has undergone
between the two images, making a clear association of features
in the two images impossible.
Whereas the 1996 image (Fig.~5 top left panel) shows two distinct shock fronts
closely together along the outflow axis, the younger of those shock fronts has apparently caught
up with the older one and by 2023, they formed one merged shock.
From its proper motion, we derive the kinematic age of
shock front 6E as 27784 d = 76 yrs and the ejection year at 1947.
Due to the morphological changes, we estimate the accuracy of the
proper motion and the kinematic age to be $\pm$10\% and the accuracy
of the ejection data as $\pm$8 yrs.

Without second epoch data, we do not know the proper motion of the other axial shock fronts
closer to the protostar
that form a string of individual shocks tracing the central axis of the flow.
Assuming, for lack of better information, 
that their proper motion is the same as that of 6E = IR2, 
their distance from the extrapolated present position of the protostar
leads to the shock front ages and ejection dates listed in Table 2.
The ages of those axial shocks close to the protostar are only of order of decades, 
similar to what was found in adaptive optics studies of other protostellar outflows, e.g., SVS 13 
\citep{Hodapp.2014.ApJ.794.169.SVS13},
and MHO3252Y3
\citep{Hodapp.2018.ApJ.864.172.MHO3252Y3}.
We compare our result to those of
\citet{Evans.2023.ApJ.943.90.B335.variability}
who interpret the start of the W2 light curve outburst as the
time of launch of a new molecular ``bullet''. This point in
time closely matches our estimate for the launch time of
the rather dim shock 2E, but the light curve does not
show a similar maximum coinciding with the launch of the
much brighter shock 3E.

We have tentatively included the feature 0E in this list, assuming that
this is the youngest shock front that is just emerging at present.
Since this object is so faint and does not have much contrast with the 
surrounding emission, we could not obtain a spectrum of it.
However, \citet{Rubinstein.2023.arXiv231207807R}
have published a JWST NIRSpec IFU spectrum of this feature, which
will be discussed in section 3.4.
We expect that with a second epoch NIRCam image expected in 2024, 
the other axial shocks close to the protostar should also have measurable proper motions.


\begin{deluxetable*}{lllllll}
\tabletypesize{\scriptsize}
\tablecaption{B335 Shock Fronts\\}
\tablewidth{0pt}
\vspace{0.5cm}
\tablehead{
\colhead{Identification} & \colhead{RA} & \colhead{Dec} & \colhead{Distance} &
\colhead{Ejection Age [d]} & \colhead{Ejection JD} & \colhead{Ejection Date}}
\startdata
HH119 JWST 0E  & 294.25380 & 7.56928 & 0.12 &  265  & 2459794 & 2022 Aug.\\
HH119 JWST 1E  & 294.25397 & 7.56927 & 0.75 & 1654  & 2458405 & 2018 Oct.\\
HH119 JWST 2E  & 294.25414 & 7.56927 & 1.34 & 2954  & 2457105 & 2015 Mar.\\
HH119 JWST 3E  & 294.25434 & 7.56928 & 2.08 & 4587  & 2455472 & 2010     \\
HH119 JWST 3W  & 294.25307 & 7.56921 &      &       &         &          \\
HH119 JWST 4E  & 294.25550 & 7.56930 & 4.65 & 10254 & 2449805 & 1995     \\
HH119 JWST 4W  & 294.25222 & 7.56904 &      &       &         &          \\
HH119 JWST 5E  & 294.25667 & 7.56887 & 10.5 & 23153 & 2436906 & 1959     \\
HH119 JWST 6E  & 294.25723 & 7.56897 & 12.6 & 27784 & 2432275 & 1947     \\
HH119 JWST 6W  & 294.24943 & 7.56784 &      &       &         &          \\
HH119 JWST 7E  & 294.26113 & 7.56870 & 27.0 & 59537 & 2400522 & 1860     \\
HH119 JWST 8E  & 294.26367 & 7.56848 & 35.6 & 78501 & 2381558 & 1808     \\
\enddata
\tablecomments{The distance in arcsec is measured from the extrapolated position of the protostar.
The ejection ages assume that all shock fronts listed here have the same proper motion as 6E}
\end{deluxetable*}

\clearpage

\subsection{Emission from the youngest shock fronts}

The marginally resolved 13 year old bright emission knot labelled 3E in Fig.\ 4 is indeed a shock front. 
One NIRCam long-wave module pixel subtends $\approx$ 10 AU at the distance of B335,
so this is the approximate linear size of the marginally resolved knots.
The axial shock front 3E is bright enough relative to the surrounding emission in the
eastern outflow lobe that
in the F460M WFSS frame in Fig.\ 2, it shows a series of emission lines 
that are identified as ro-vibrational transitions of CO 
in the extracted spectrum in Fig.\ 6 (top panel). 
CO bandhead emission is frequently seen in the spectra 
of protostars and young star themselves, a recent example being
the work on the B335 protostar
and its disk, seen in scattered light in
the innermost parts of the outflow cavity by
\citet{Rubinstein.2023.arXiv231207807R}. Their NIRSpec IFU pointing
contained object 0E in our nomenclature, which they name B335-E, and the reflection nebulosity
immediately west of the protostar that they call B335-W and we have not individually labelled.
They detect strong emission from the CO ro-vibrational transitions in the range from 4.4 - 5.2 $\mu$m.
While \citet{Rubinstein.2023.arXiv231207807R} work under the assumption that this emission
is scattered light from the protostar and its disk, their result leaves the question open
whether 0E (their B335-E) is actually an emerging shock front.
Recently, JWST observations of HH211 by \citet{Ray.2023.Natur.622.48.HH211}
have shown CO emission around 4.6 $\mu$m not only from scattered light from the
immediate vicinity of the protostar, but also in the bow shocks of that outflow.

\begin{figure}[h]
\begin{center}
	\includegraphics[angle=0.,scale=0.40]{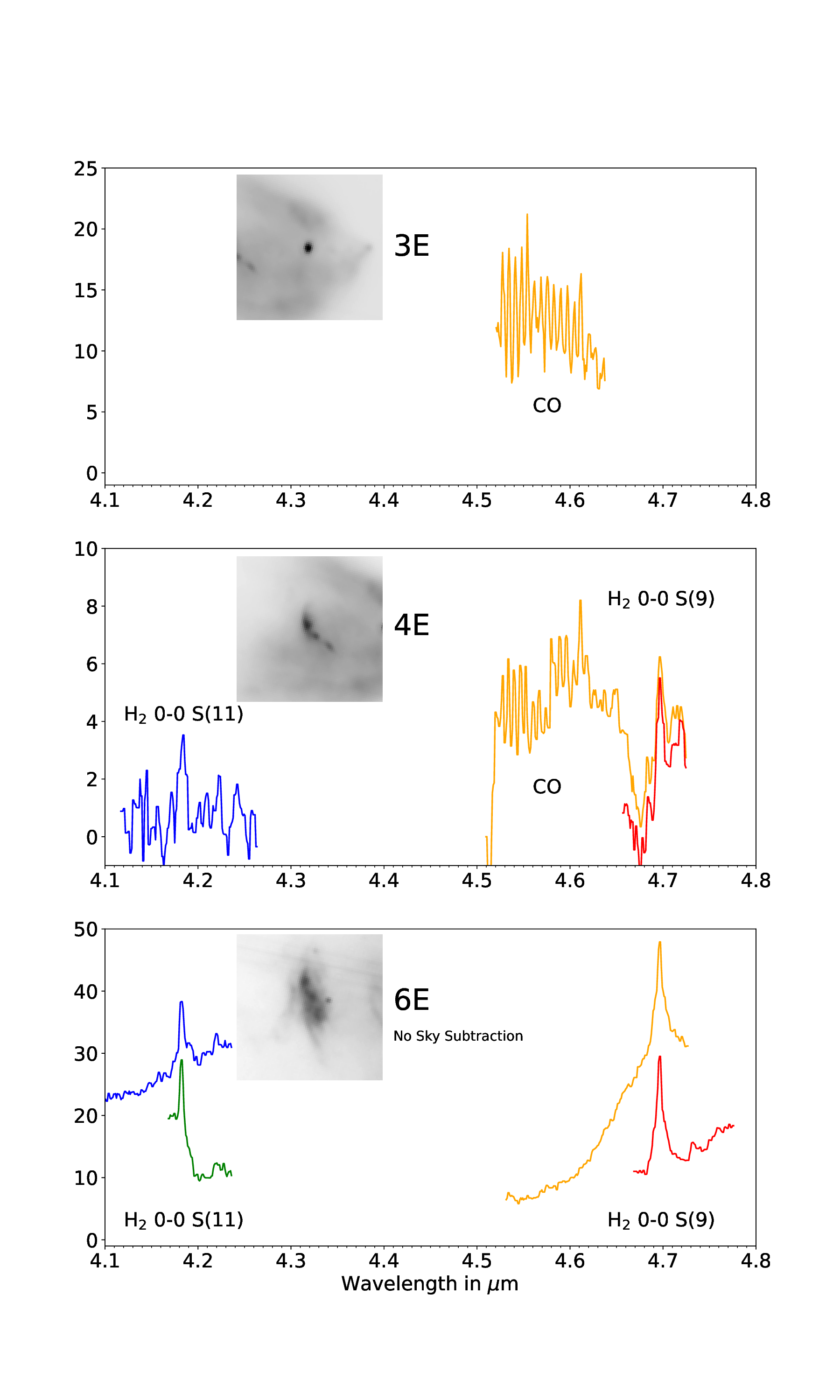}
	\caption{
        Extracted WFSS spectra of the three brightest (3E, 4E, 6E)  of the young axial shock fronts in the eastern lobe
of the B335 outflow.
For orientation, small postage stamp cutouts of the direct image of each of the three shock
fronts are included with North up and East left. The WFSS spectra extracted in each filter are
color-coded: F410M (blue), F430M (green), F460M (orange), and F480M (red). All spectra shown here
were divided by the transmission profile of filter, and spectral regions recorded with less than 10\% filter
transmission were excluded.
The spectra of shock regions 3E and 4E are sky-subtracted.
For the most extended of these three shock fronts, 6E, we did not subtract the signal from
nearby sky positions, since there was no ''sky'' without H$_2$ emission available.
Also, this spectrum records the H$_2$ emission of the fainter shock region 5E that, due
to the degeneracy between spatial and spectral information in slitless spectroscopy data,
are recorded at the wrong, longer wavelength.
        }
	\end{center}
\end{figure}

The bright, extended scattered light in the eastern outflow lobe
shows deep absorption in the WFSS frame through the F430M filter in Fig.~2. 
This absorption feature, broadened by the extended flux
distribution of the scattered light in the outflow cavity, is indicative
of absorption by CO$_2$ ice in the B335 molecular core.
Our WFSS spectroscopy of background stars behind B335, the original
purpose of our program 1187, shows saturated CO$_2$ ice absorption
near the center of the B335 core, which will be published in a 
separate paper.

The next of the inner axial shock fronts is the more extended 28-year-old 4E that has a clear
bow shock morphology. We show the extracted spectrum in Fig.\ 6 (middle panel).
The emission from CO is clearly detected, and the emission lines of
H$_2$ are more prominent relative to CO than in the case of the younger,
more compact 3E. 
The third spectrum, the bottom panel in Fig.\ 6, is the extracted spectrum
of the 76-year-old axial shock 6E and is dominated by the two H$_2$ emission lines that fall
in the transmission bands of the filters used here. We do not detect
CO emission in this shock.
Similarly, the more distant older shock fronts are emitting in the 0-0 S(9) (4.6946 $\mu$m) 
and S(11) (4.1811 $\mu$m)
lines of H$_2$.
The H$_2$ 0-0 S(10) line is just outside the bandpass of filter F430M and therefore not recorded.

Assuming that the morphology, extent, and emission line characteristics
of these three shocks represent a pattern, we conclude that typical shocks
emerge from the protostar as fairly small ($\approx$ 10 AU) clumps of material, expand over 
the course of a few decades, and change their density and temperature so
that CO ro-vibrational lines are no longer excited.
In the recent past, the B335 protostar appears to have undergone velocity changes in
its molecular jet, leading to the production of internal shock fronts
every 3 to 5 years.
The apparently much higher frequency of shock ejection 
in recent decades compared to the larger separation and therefore lower shock
frequency in the older shocks. This can be understood if shock fronts are initially
ejected at a higher than average velocity and over time catch up with slower moving
outflow material ahead of them. This scenario of individual internal work surfaces in
a jet propagating into a dense medium was developed and modeled by 
\citet{Raga.1990.ApJ.364.601.jets.var}, applied to HH47, and
further developed for jets of non-uniform density by \citet{Raga.1992.ApJ.386.222.nonuniform.jet.var},
applied to the case of HH34.
Their scenario is similar to what we actually observed for shock front 6E,
where two closely spaced shock fronts appear to have merged over the course of
the past 27 years.
We explain the fact that only a few strong shocks remain detectable after centuries
by the combined effect of shock fronts expanding sideways and of shock fronts merging. 

\section{Summary and Conclusions}

While many of the more distant and less obscured shock fronts have been studied before, 
the deep JWST images in the NIRCam long wavelength channel presented here
show the full bipolar nebula associated with the B335 protostar. 
With much higher spatial resolution than older $Spitzer$ images, 
symmetric sets of shock fronts were found in the outflow lobes 
that indicate the repeated ejection of fast-moving gas whose
differential velocities  
lead to the formation of these internal, relatively low excitation shock fronts. 
The kinematic age of the closest shock fronts from the protostar is of the order of a few years, 
if we assume the same proper motion as the closest shock with measured proper motion (6E). 
The youngest shock front with a spectrum (3E) is dominated by emission lines of gaseous CO. 
The next older shock front (4E) still shows CO emission, but with diminished strength, 
while more distant and older shock fronts (6E) do not show CO, but are dominated by H$_2$ emission lines.
The string of shock fronts emerging from the B335 protostar is indicative 
of repeated ejection events, probably due to some form of accretion episodicity. 
Among the older shock fronts at larger distances from the protostar, 
we note a distinction into two classes of proper motion, based on the measurements by 
\citet{Galfalk.2007.A&A.475.281.B335.outflows}.
The low proper motion peripheral
shocks in our images have their bow shocks pointing towards the protostar, 
indicative of shocks on a nearly stationary clump of denser material where
the outflowing gas flows around the obstacle.
By contrast, all high proper motion axial shock fronts have their bow shocks 
pointing away from the protostar, 
indicating shocks of the outflow wind against near stationary and uniform ambient gas.
For the shock front 6E, we document the apparent merger of two distinct shock front
seen in 1996 into one shock front in 2023.

\begin{acknowledgments}
This project was supported by NASA
through the JWST /NIRCam project, 
contract number NAS5-02105 (M. Rieke, University of Arizona, PI).
The data were obtained under GTO program 1187,
downloaded from the Mikulski Archive for Space Telescopes
(MAST) at the Space Telescope Science Institute
and are available at MAST: \dataset[10.17909/qa3f-5703]{\doi{10.17909/qa3f-5703}}.
The first epoch image of B335 was obtained at the W. M.
Keck Observatory, which is operated as a scientific partnership
among the California Institute of Technology, the University
of California and NASA. The Observatory was made possible
by the generous financial support of the W. M. Keck Foundation.
This work is based
in part on observations made with the Spitzer Space Telescope, which was operated by the Jet
Propulsion Laboratory, California Institute of Technology under a contract with NASA.
D.J.\ is supported by NRC Canada and by an NSERC Discovery Grant.
We thank Bo Reipurth for helpful discussions and comments and
Thomas Dutkiewicz for help with the mining of the DOIs.
We thank the referee for constructive comments that helped improve the paper.
\end{acknowledgments}

\vspace{5mm}
\facilities{JWST, Keck:I}



\end{document}